\documentclass{ifacconf}

\usepackage{natbib}
\usepackage{cuted}
\usepackage{lipsum}
\usepackage{graphicx}
\usepackage{url}
\usepackage{mathtools}
\usepackage{tikz}
\usepackage[americanresistors,americaninductors]{circuitikz}
\usetikzlibrary{shapes,arrows}
\usetikzlibrary{automata}
\usepackage{amsmath}
\usepackage{verbatim}
\usepackage{listings}
\usepackage{amssymb}
\usepackage{mathrsfs}
\usepackage{enumerate}
\allowdisplaybreaks

\usepackage{tikz}
\usetikzlibrary{calc,patterns,decorations.pathmorphing,decorations.markings}
\usepackage{pgfplots}

\usetikzlibrary{dsp,fit, shapes,arrows, calc, chains, positioning, quotes, shadows}
\tikzstyle{block} = [draw, rectangle, 
    minimum height=3em, minimum width=4em,align=center, rounded corners=0.1cm, fill=gray!0!white]
\tikzstyle{cblock} = [draw, rectangle, 
    minimum height=3em, minimum width=4em,align=center, rounded corners=0.1cm, fill=gray!25!white]
\tikzstyle{circ} = [draw, circle, 
   node distance=3em ,align=center, fill=gray!0!white]
\tikzstyle{ccirc} = [draw, circle, 
   node distance=3em ,align=center, fill=gray!25!white]
\tikzstyle{blueblock} = [draw, rectangle, 
    minimum height=3em, minimum width=4em,align=center, color = blue]
\tikzstyle{blckblock} = [draw, rectangle, 
    minimum height=3em, minimum width=4em,align=center, color = black]    
\tikzstyle{hblock} = [draw, rectangle, 
    minimum height=15em, minimum width=3em,align=center]    
\tikzstyle{eblock} = [draw, rectangle, 
    minimum height=3em, minimum width=4em,align=center, rounded corners=0.1cm, fill=white, color=white]
\tikzstyle{outerblock} = [draw, rectangle, dashed, inner sep = .85em]
\tikzstyle{sum} = [draw, circle, node distance=1cm]
\tikzstyle{var} = [draw, rectangle, minimum height = 3em, minimum width=0em]
\tikzstyle{pinstyle} = [pin edge={to-,thin,black}]
\tikzstyle{input} = [coordinate]
\tikzstyle{output} = [coordinate]

{\theorembodyfont{\slshape}\newtheorem{theorem}{Theorem}[section]}
{\theorembodyfont{\upshape}\newtheorem{proposition}[theorem]{Proposition}}  %
{\theorembodyfont{\slshape}}
{\theorembodyfont{\slshape}}           %
{\theorembodyfont{\slshape}}
{\theorembodyfont{\upshape}\newtheorem{corollary}[theorem]{Corollary}}  %
{\theorembodyfont{\upshape}\newtheorem{definition}[theorem]{Definition}}
{\theorembodyfont{\upshape}}
{\theorembodyfont{\upshape}\newtheorem{remark}[theorem]{Remark}}
{\theorembodyfont{\upshape}}
{\theorembodyfont{\upshape}\newtheorem{example}[theorem]{Example}}
{\theorembodyfont{\upshape}\newtheorem{algorithmIF}[theorem]{Algorithm}}

\begin{document}
\begin{frontmatter}
\title{Scalable distributed and decentralized $\mathscr{H}_2$ controller synthesis for interconnected linear discrete-time systems}\thanks{This work has received funding from the European Research Council (ERC), Advanced Research Grant SYSDYNET, under the European Union’s Horizon 2020 research and innovation programme (Grant Agreement No. 694504).}

\author{Tom R.V. Steentjes, Mircea Lazar, Paul M.J. Van den Hof}

\begin{abstract}
The current limitation in the synthesis of distributed $\mathscr{H}_2$ controllers for linear interconnected systems is scalability due to non-convex or unstructured synthesis conditions. In this paper we develop convex and structured conditions for the existence of a distributed $\mathscr{H}_2$ controller for discrete-time interconnected systems with an interconnection structure that corresponds to an arbitrary graph. Neutral interconnections and a storage function with a block-diagonal structure are utilized to attain coupling conditions that are of a considerably lower computational complexity compared to the corresponding centralized $\mathscr{H}_2$ controller synthesis problem. Additionally, the developed conditions are adapted for the corresponding decentralized $\mathscr{H}_2$ controller synthesis problem with fixed supply functions for the interconnections. The effectiveness and scalability of the developed distributed $\mathscr{H}_2$ controller synthesis method is demonstrated for small- to large-scale oscillator networks on a cycle graph.
\end{abstract}

\end{frontmatter}

\section{Introduction}
Control of interconnected systems is relevant to a wide area of applications in smart grids, communication networks, irrigation networks and chemical plant networks, fueled by the digital industrial revolution, see e.g. \citep{lunze92} and \citep{bullo2018}. Distributed control is preferred for such systems due to its scalable implementation and it has been a major research topic in recent years for several control objectives, including $\mathscr{H}_2$ and $\mathscr{H}_\infty$ performance criteria.

For continuous-time systems, sufficient conditions for the existence of a controller that admits the same interconnection structure as the plant and that achieves unit $\mathscr{H}_\infty$ performance were developed by \cite{langbortetal2004}. The basis for these sufficient conditions is laid by dissipativity theory, introduced by \cite{willems72}, which is also the cornerstone for this work. \cite{horssen2016} presented a discrete-time analogue of the work in \cite{langbortetal2004} with additional robust stability and robust $\mathscr{H}_\infty$ performance guarantees. For both the continuous- and discrete-time distributed $\mathscr{H}_\infty$ control problems, the conditions can be stated as linear matrix inequalities (LMIs) \citep{langbortetal2004}, \citep{horssen2016}.

\cite{eilbrecht2017} provided an approach to solve the discrete-time $\mathscr{H}_2$ output-feedback problem for interconnected systems, by minimizing a linear combination of the closed-loop system's $\mathscr{H}_2$ norm and a cost related to the sparsity of the controller matrices. However, this approach yields a non-convex problem in general. \cite{vamsi2016} solved the discrete-time $\mathscr{H}_2$ problem for a `strictly causal' network, via the search for an unstructured controller and a subsequent transformation into a structured one. The structure of systems interconnected over one spatial dimension was exploited by \cite{rice2010} for the efficient design of $\mathscr{H}_2$ controllers interconnected in a string. The distributed $\mathscr{H}_2$ controller synthesis for continuous-time systems with arbitrary interconnection topology was recently considered by \cite{chen2019}. Unlike the $\mathscr{H}_\infty$ case, however, the feasibility problem for the distributed $\mathscr{H}_2$ controller existence in \citep{chen2019} is not convex, but amounts to solving a bilinear optimization problem.

The $\mathscr{H}_2$ norm has a particularly interesting interpretation in the field of data-driven modelling of interconnected systems, where stochastic assumptions on disturbance signals are key \citep{vandenhofetal2013}. This is due to the fact that the $\mathscr{H}_2$ norm equals the asymptotic output variance for a white noise excitation \citep{schererweilandLMI}. The trend for data-driven modelling of interconnected systems asks for accompanying distributed controller design methods that apply to discrete-time systems affected by stochastic disturbance signals. However, the current approaches to distributed $\mathscr{H}_2$ control, reviewed above, do not facilitate the controller synthesis for arbitrarily-structured large-scale systems, due to non-convex or unstructured synthesis conditions, or due to restrictions to systems that are spatially distributed in one dimension. Hence, it is of interest to develop scalable (convex) conditions for the synthesis of distributed $\mathscr{H}_2$ controllers for systems with a general interconnection structure.

In this paper, we therefore develop sufficient conditions for the existence of a distributed $\mathscr{H}_2$ controller for a discrete-time system with an arbitrary interconnection structure, by adopting the fundamental approach to distributed controller synthesis of \cite{langbortetal2004}. Analogous to distributed $\mathscr{H}_2$ controller synthesis for linear continuous-time systems \citep{chen2019}, the conditions are principally not convex, which is induced by a number of scalar terms that are nonlinear w.r.t. the optimization variables, equal to the number of subsystems. However, we show that the resulting conditions are equivalent to alternative \emph{convex} conditions stated as LMIs, with no reduction in generality or scalability. Additionally, we adapt our conditions for the existence of a decentralized $\mathscr{H}_2$ controller, by imposing a dissipative property for controlled subsystems with respect to the interconnection channels, such as passivity, to facilitate the implementation in applications where communication between controllers is not practical.

This paper is organized as follows: in Section~2 we give a description of the interconnected system and analysis conditions. These conditions are used to provide convex existence conditions and a construction procedure for distributed and decentralized $\mathscr{H}_2$ controllers in Section~3. In Section~4, we present a numerical example where distributed $\mathscr{H}_2$ controller synthesis is illustrated for an oscillator network with a cyclic interconnection structure and compared with a centralized $\mathscr{H}_2$ controller in terms of scalability. Conclusions are summarized in Section~5.

\subsection*{Basic nomenclature}
The integers are denoted by $\mathbb{Z}$. Given $a\in \mathbb{Z}$, $b\in \mathbb{Z}$ such that $a<b$, we denote $\mathbb{Z}_{[a:b]}:=\left\{a,a+1,\dots,b-1,b\right\}$. Let $I_n\in\mathbb{R}^{n\times n}$, or simply $I$, denote the identity matrix. The operator $\operatorname{col}(\cdot)$ stacks its arguments in a column vector. The block diagonal matrix $\operatorname{diag}(X_1,\dots,X_m)$ has matrices $X_i$, $i\in\mathbb{N}_{[1:m]}$, in its block diagonal entries. For $S\subseteq \mathbb{Z}$, the block diagonal matrix $\operatorname{diag}_{i\in S} X_i$ has matrices $X_i$, $i\in S$, in its block diagonal entries. The image of a matrix $A\in\mathbb{R}^{m\times n}$ is $\operatorname{im} A:=\{Ax\,|\, x\in\mathbb{R}^n\}$. For a real symmetric matrix $X$, $X\succ 0$ denotes that $X$ is positive definite.

\section{Preliminaries}
Let the structure of an interconnected system be given by a graph $G=(V,E)$, where $V$ is the vertex set of cardinality $L$ and $E\subseteq V\times V$ is the edge set. Each vertex $v_i\in V$, corresponds to a discrete-time system $\mathcal{P}_i$. An edge $(v_i,v_j)\in E$ exists if subsystems $\mathcal{P}_i$ and $\mathcal{P}_j$ are interconnected. For ease of presentation, self-connections are excluded for all subsystems $\mathcal{P}_i$, $i\in\mathbb{Z}_{[1:L]}$.

Each subsystem $\mathcal{P}_i$ is assumed to admit a state-space representation
\begin{align} \label{eq:subsys}
\begin{pmatrix}
x_i(k+1)\\ o_i(k)\\ z_i(k)
\end{pmatrix}=\begin{pmatrix}
A_i^{\mathrm{T}\mathrm{T}} & A_i^{\mathrm{T}\mathrm{S}} & B_i^{\mathrm{T}d} & B_i^{\mathrm{T}u}\\ A_i^{\mathrm{S}\mathrm{T}} & A_i^{\mathrm{S}\mathrm{S}} & B_i^{\mathrm{S}d} & B_i^{\mathrm{S}u}\\ C_i^{z\mathrm{T}} & C_i^{z\mathrm{S}} & D_i^{zd} & D_i^{zu}
\end{pmatrix}\begin{pmatrix}
x_i(k)\\ s_i(k)\\ d_i(k)
\end{pmatrix},
\end{align}
where $x_i:\mathbb{Z}\to \mathbb{R}^{k_{i}}$ is the subsystem's state, $o_i:\mathbb{Z}\to\mathbb{R}^{n_i}$ and $s_i:\mathbb{Z}\to\mathbb{R}^{n_i}$ are the outgoing and incoming interconnection variables, and $z_i:\mathbb{Z}\to\mathbb{R}^{q_i}$ and $d_i:\mathbb{Z}\to\mathbb{R}^{f_i}$ are the performance output and disturbance input, respectively.

We write the interconnection signals $s_i$ and $o_i$ as $s_i=\operatorname{col}(s_{i1},s_{i2},\dots,s_{iL})$ and $o_i=\operatorname{col}(o_{i1},o_{i2},\dots,o_{iL})$ so that $(s_{ij},o_{ij})$ denotes the interconnection channel between subsystem $\mathcal{P}_i$ and subsystem $\mathcal{P}_j$. For the ease of the interconnection definition, we assume, without loss of generality \cite{langbortetal2004}, that $o_{ij}$, $s_{ij}$, $o_{ji}$ and $s_{ji}$ are all elements of $\mathbb{R}^{n_{ij}}$, $n_{ij}\geq 0$. The interconnection between system $\mathcal{P}_i$ and $\mathcal{P}_j$ is defined through the interconnection equation
\begin{align} \label{eq:int}
\begin{pmatrix}
o_{ij}(k)\\s_{ij}(k)
\end{pmatrix}=\begin{pmatrix}
s_{ji}(k)\\o_{ji}(k)
\end{pmatrix},\quad \forall k\in\mathbb{Z}.
\end{align}
Hence, $\mathcal{P}_i$ and $\mathcal{P}_j$ are interconnected if and only if $n_{ij}>0$, if and only if $(v_i,v_j)\in E$.

The interconnected system can be compactly represented by
\begin{align*}
\begin{pmatrix}
x(k+1)\\ o(k)\\ z(k)
\end{pmatrix}=\begin{pmatrix}
A^{\mathrm{TT}} & A^{\mathrm{TS}} & B^\mathrm{T}\\ A^{\mathrm{ST}} & A^{\mathrm{SS}} & B^\mathrm{S}\\ C^\mathrm{T} & C^\mathrm{S} & D
\end{pmatrix}\begin{pmatrix}
x(k)\\ s(k)\\ d(k)
\end{pmatrix},
\end{align*}
with corresponding definitions for the system matrices and signals, cf. \citep{steentjes2020}, and the interconnection $o=\Delta s$, with the matrix $\Delta$ defined by aggregating \eqref{eq:int} for all corresponding index pairs. Elimination of the interconnection variables $s$ and $o$ yields a state-space representation
\begin{align} \label{eq:PIss}
\mathcal{P}_\mathcal{I}:\quad \begin{pmatrix}
x(k+1)\\ z(k)
\end{pmatrix}=\begin{pmatrix}
A_\mathcal{I} & B_\mathcal{I}\\ C_\mathcal{I} & D_\mathcal{I}
\end{pmatrix}\begin{pmatrix}
x(k)\\ d(k)
\end{pmatrix}
\end{align}
where
\small
\begin{align*}
\begin{pmatrix}
A_\mathcal{I} & B_\mathcal{I}\\ C_\mathcal{I} & D_\mathcal{I}
\end{pmatrix}&:=\begin{pmatrix}
A^{\mathrm{TT}} & B^\mathrm{T}\\ C^\mathrm{T} & D
\end{pmatrix}+\begin{pmatrix}
A^{\mathrm{TS}}\\ C^\mathrm{S}
\end{pmatrix}(\Delta-A^{\mathrm{SS}})^{-1}\begin{pmatrix}
A^{\mathrm{ST}} & B^\mathrm{S}
\end{pmatrix}.
\end{align*}
\normalsize
Consider the interconnection variable subspaces \citep{langbortetal2004}
\begin{align*}
\mathcal{S}_\mathcal{I}&:=\{(o,s)\in\mathbb{R}^{2n}\,|\, o=\Delta s\} \text{ and }\\
\mathcal{S}_\mathcal{B}&:=\{(o,s)\in\mathbb{R}^{2n}\,|\, \operatorname{col}(o_i,s_i)\in\operatorname{im} \operatorname{col} (A_i^\mathrm{SS},I),\,\! i\in\mathbb{Z}_{[1:L]}\}.
\end{align*}
\begin{definition}
An interconnected system described by \eqref{eq:subsys} and \eqref{eq:int} is said to be well-posed if $\mathcal{S}_\mathcal{I}\cap \mathcal{S}_\mathcal{B}=\{0\}$.
\end{definition}
\begin{definition}
A well-posed interconnected system is said to be asymptotically stable (AS) if the roots of $\det(zI-A_\mathcal{I})$ are inside the unit circle on the complex plane.
\end{definition}
\begin{definition}
The $\mathscr{H}_2$ norm of a well-posed and AS interconnected system with a transfer function $T(z):=C_\mathcal{I}(zI-A_\mathcal{I})^{-1}B_\mathcal{I}+D_\mathcal{I}$ is defined by
\begin{align*}
\|\mathcal{P}_\mathcal{I}\|_{\mathscr{H}_2}:=\left(\frac{1}{2\pi}\operatorname{trace}\int_{-\pi}^\pi T^{*}(e^{i\omega})T(e^{i\omega})\,\mathrm{d}\omega\right)^{\frac{1}{2}}.
\end{align*}
\end{definition}

\subsection{Interconnected-system analysis} \label{sec:uncsys}
As a basis for the analysis of the interconnected system and the synthesis of distributed controllers, we employ the theory of dissipative dynamical systems \citep{willems72}.
\begin{definition}
Subsystem $\mathcal{P}_i$ is said to be dissipative with respect to the supply function $\sigma_i:\mathcal{S}_i\times \mathcal{O}_i\times \mathcal{D}_i\times\mathcal{Z}_i\to \mathbb{R}$, if there exists a non-negative storage function $V_i:\mathcal{X}_i\to\mathbb{R}_{\geq 0}$, so that for all $t\in\mathbb{Z}_{\geq 0}$ the inequality
\begin{align*}
V_i(x_i(t))-V_i(x_i(0))\leq \sum_{k=0}^{t-1}\sigma_i(s_i(k),o_i(k),d_i(k),z_i(k))
\end{align*}
holds for all trajectories $(x_i,s_i,o_i,d_i,z_i)$ of \eqref{eq:subsys}.
\end{definition}

We consider the class of quadratic storage functions:
\begin{align*}
V_i(x_i):= x_i^\top X_i x_i,\quad i\in\mathbb{Z}_{[1:L]},
\end{align*}
with $X_i\succ 0$. Supply functions are restricted to be quadratic functions of the form
\begin{align*}
\sigma_i(s_i,o_i,d_i,z_i):=\sigma_i^{\mathrm{int}}(s_i,o_i)+\sigma_i^{\mathrm{ext}}(d_i,z_i),\quad i\in\mathbb{Z}_{[1:L]},
\end{align*}
with `internal' supply functions
\begin{align*}
\sigma_i^{\mathrm{int}}(s_i,o_i)&:=\sum_{j=1}^L \sigma_{ij}(s_{ij},o_{ij}),\\
\sigma_{ij}(s_{ij},o_{ij})&:=\begin{pmatrix}
o_{ij}\\s_{ij}
\end{pmatrix}^\top X_{ij}\begin{pmatrix}
o_{ij}\\ s_{ij}
\end{pmatrix},
\end{align*}
where $X_{ij}$ is a real symmetric matrix, and `external' supply functions
\begin{align*}
\sigma_i^{\mathrm{ext}}(d_i,z_i)&:=\rho_id_i^\top d_i -z_i^\top z_i,
\end{align*}
where $\rho_i>0$. For any pair $(i,j)\in\mathbb{Z}_{[1:L]}^2$, $i\neq j$, the interconnection between subsystem $\mathcal{P}_i$ and subsystem $\mathcal{P}_j$ is said to be neutral if the internal supply functions satisfy \citep{schererweilandLMI}
\begin{align} \label{eq:neutral}
0=\sigma_{ij}(s_{ij},o_{ij})+\sigma_{ji}(s_{ji},o_{ji}).
\end{align}
One can interpret a neutral interconnection as a lossless one; no `energy' is dissipated or supplied through the interconnection channel \citep{willems72}. The neutrality condition \eqref{eq:neutral} is equivalent with
\begin{align*}
0=X_{ij}+\begin{pmatrix}
0 & I\\ I & 0
\end{pmatrix}X_{ji}\begin{pmatrix}
0 & I\\ I & 0
\end{pmatrix}.
\end{align*}

The following result provides sufficient conditions for well-posedness, stability and bounding the $\mathscr{H}_2$ norm of the interconnected system, and provides a discrete-time counterpart of the continuous-time result \cite[Theorem~1]{chen2019}. Define the matrix
\begin{align} \label{eq:Ti}
T_i:=\left(\begin{array}{ccc}
I & 0 & 0\\ A_i^{\mathrm{T}\mathrm{T}} & A_i^{\mathrm{T}\mathrm{S}}  & B_i^\mathrm{Td} \\ \hline
A_i^{\mathrm{S}\mathrm{T}} & A_i^{\mathrm{S}\mathrm{S}} & B_i^\mathrm{Sd} \\ 0 & I & 0\\ \hline
C_i^\mathrm{zT}  & C_i^\mathrm{zS}  & D_i^{zd} \\
0 & 0 & I
\end{array}\right).
\end{align}

\begin{proposition} \label{prop:an}
The interconnected system $\mathcal{P}_\mathcal{I}$ is well-posed, AS and $\|\mathcal{P}_\mathcal{I}\|_{\mathscr{H}_2}<\gamma$, if $B_i^{\mathrm{S}d}=0$ for all $i\in\mathbb{Z}_{[1:L]}$ and there exist positive-definite $X_i\in\mathbb{R}^{k_{i}\times k_{i}}$, $\rho_i>0$, symmetric $X_{ij}^{11}\in\mathbb{R}^{n_{ij}\times n_{ij}}$, $(i,j)\in\mathbb{Z}_{[1:L]}^2$, and $X_{ij}^{12}\in\mathbb{R}^{n_{ij}\times n_{ij}}$, $(i,j)\in\mathbb{Z}_{[1:L]}^2$, $i>j$, with
\begin{align}
&T_i^\top \left(\begin{array}{cc|cc|cc}
-X_i & 0 & 0 & 0 & 0 & 0\\
0 & X_i & 0 & 0& 0 & 0\\ \hline
0 & 0 & Z_i^{11}& Z_i^{12} & 0 & 0\\ 0 & 0 & (Z_i^{12})^\top & Z_i^{22} & 0& 0\\ \hline
0 & 0 & 0 & 0 & I & 0\\ 0 & 0 & 0 & 0 & 0 & -\rho_i I
\end{array}\right)T_i\prec 0,\label{eq:cross}\\
& \sum_{i=1}^L \operatorname{trace} \left( (B_i^{\mathrm{T}d})^\top X_i B_i^{\mathrm{T}d}+(D_i^{zd})^\top D_i^{zd}\right)<\gamma^2,\label{eq:cross2}
\end{align}
where
\begin{align*}
Z_i^{11} &:=-\operatorname*{diag}_{j\in\mathbb{Z}_{[1:L]}} X_{ij}^{11},\ Z_i^{22} :=\operatorname*{diag}_{j\in\mathbb{Z}_{[1:L]}} X_{ji}^{11},\\
Z_i^{12} &:=\operatorname{diag}\left(-\operatorname*{diag}_{j\in\mathbb{Z}_{[1:i]}} X_{ij}^{12},\operatorname*{diag}_{j\in\mathbb{Z}_{[i+1:L]}} (X_{ji}^{12})^\top\right).
\end{align*}
\end{proposition}
\begin{pf}
Well-posedness is identically defined for continuous-time systems \citep{langbortetal2004}, hence we refer the reader to \citep[Theorem~1]{langbortetal2004} for the proof of well-posedness, since \eqref{eq:cross} implies the condition used therein for well-posedness.

Let \eqref{eq:cross} and \eqref{eq:cross2} be true. We define the candidate local storage functions $V_i(x_i):=x_i^\top X_ix_i$ and the candidate global storage function $V(x):=\sum_{i=1}^{L} V_i(x_i)$. Multiplication of inequality \eqref{eq:cross} from the right and from the left with $\operatorname{col}(x_i(k),s_i(k),d_i(k))$ and its transpose yields
\begin{align*}
0&>x_i^\top(k+1)X_ix_i(k+1)-x_i^\top(k)X_ix_i(k)\\
&\quad+\begin{pmatrix}
o_i(k)\\ s_i(k)
\end{pmatrix}^\top \begin{pmatrix}
Z_i^{11} & Z_i^{12}\\ (Z_i^{12})^\top & Z_i^{22}
\end{pmatrix} \begin{pmatrix}
o_i(k)\\ s_i(k)
\end{pmatrix}\\
&\quad +z_i^\top(k) z_i(k)-\rho_id_i^\top(k) d_i(k)\\
&=V_i(x(k+1))-V_i(x(k))-\sigma_i^\mathrm{int}(s_i(k),o_i(k))\\
&\quad-\sigma_i^\mathrm{ext}(d_i(k),z_i(k)),
\end{align*}
i.e., system $\mathcal{P}_i$ is dissipative with respect to the supply function $\sigma_i$. Summing the latter inequality over $i$ yields
\begin{align*}
V(x(k+1))-V(x(k))<\sum_{i=1}^L \sigma_i^\mathrm{int}+\sigma_i^\mathrm{ext}.
\end{align*}
From the neutrality condition \eqref{eq:neutral}, we observe that $\sum_{i=1}^L \sigma_i^\mathrm{int}=0$, and thus
\begin{align} \label{eq:lfineq}
V(x(k+1))-V(x(k))<\sum_{i=1}^L\sigma_i^\mathrm{ext}.
\end{align}
To prove stability, consider the case that $d(k)=0$. Then
\begin{align*}
V(x(k+1))-V(x(k))<-\sum_{i=1}^Lz_i^\top(k)z_i(k)\leq 0.
\end{align*}
Therefore, $V$ is a Lyapunov function for the interconnected system $\mathcal{P}_\mathcal{I}$ with $d(k)=0$, from which we conclude asymptotic stability of the interconnected system \cite[Corollary~1.2]{kalman1960}.

Next, we prove $\mathscr{H}_2$ performance for $\mathcal{P}_\mathcal{I}$. From \eqref{eq:PIss} and \eqref{eq:lfineq}, it follows that for all $(x,d)$
\begin{align*}
&\begin{pmatrix}
x\\ d
\end{pmatrix}^\top\begin{pmatrix}
I & 0\\ A_\mathcal{I} & B_\mathcal{I}
\end{pmatrix}^\top\begin{pmatrix}
-X_\mathcal{I} & 0\\ 0 & X_\mathcal{I}
\end{pmatrix}\begin{pmatrix}
I & 0\\ A_\mathcal{I} & B_\mathcal{I}
\end{pmatrix}\begin{pmatrix}
x\\ d
\end{pmatrix}\\
&\quad < -\begin{pmatrix}
x\\ d
\end{pmatrix}^\top\begin{pmatrix}
C_\mathcal{I} & D_\mathcal{I}\\ 0 & I
\end{pmatrix}^\top \begin{pmatrix}
I & 0\\ 0 & -P
\end{pmatrix}\begin{pmatrix}
C_\mathcal{I} & D_\mathcal{I}\\ 0 & I
\end{pmatrix}\begin{pmatrix}
x\\ d
\end{pmatrix},
\end{align*}
with $X_\mathcal{I}:=\operatorname{diag}_{i\in\mathbb{Z}_{[1:L]}} X_i$ and $P:=\operatorname{diag}_{i\in\mathbb{Z}_{[1:L]}} \rho_iI$. Hence
\begin{align*}
\begin{pmatrix}
A_\mathcal{I}^\top X_\mathcal{I}A_\mathcal{I}-X_\mathcal{I} +C_\mathcal{I}^\top C_\mathcal{I} & A_\mathcal{I}^\top X_\mathcal{I}B_\mathcal{I}+C_\mathcal{I}^\top D_\mathcal{I}\\ B_\mathcal{I}^\top X_\mathcal{I}A_\mathcal{I}+D_\mathcal{I}^\top C_\mathcal{I} & B_\mathcal{I}^\top X_\mathcal{I}B_\mathcal{I}+D_\mathcal{I}^\top D_\mathcal{I}-P
\end{pmatrix}\prec 0,
\end{align*}
which implies
\begin{align}
A_\mathcal{I}^\top X_\mathcal{I}A_\mathcal{I}-X_\mathcal{I} +C_\mathcal{I}^\top C_\mathcal{I} \prec 0. \label{eq:proofH21}
\end{align}
Since $B_i^{\mathrm{S}d}=0$ for all $i\in\mathbb{Z}_{[1:L]}$, we have
\begin{align}
&\operatorname{trace} \left(B_\mathcal{I}^\top X_\mathcal{I}B_\mathcal{I}+D_\mathcal{I}^\top D_\mathcal{I}\right)\nonumber\\
&\quad =\operatorname{trace} \left(\sum_{i=1}^L (B_i^{\mathrm{T}d})^\top X_iB_i^{\mathrm{T}d}+(D_i^{zd})^\top D_i^{zd}\right)\nonumber\\
&\quad =\sum_{i=1}^L \operatorname{trace} \left( (B_i^{\mathrm{T}d})^\top X_iB_i^{\mathrm{T}d}+(D_i^{zd})^\top D_i^{zd}\right)<\gamma^2. \label{eq:proofH22}
\end{align}
Finally, \eqref{eq:proofH21}, \eqref{eq:proofH22} and $X_\mathcal{I}\succ 0$ imply $\|\mathcal{P}_\mathcal{I}\|_{\mathscr{H}_2}<\gamma$ by \citep[Proposition~II.1]{steentjes2020}, which completes the proof.\hfill $\blacksquare$
\end{pf}

We illustrate the analysis conditions in Proposition~\ref{prop:an} by a simple example.
\begin{example}
Consider two identical scalar subsystems described by
\begin{align*}
x_i(k+1)=\frac{1}{2}x_i(k)+\frac{1}{10}s_i(k)+d_i(k),\quad i=1,2,\ k\in\mathbb{Z},
\end{align*}
and $z_i(k)=o_i(k)=x_i(k)$, with interconnection constraints $s_1(k)=o_2(k)$, $s_2(k)=o_1(k)$. It is easily verified that LMI \eqref{eq:cross} holds for $i=1,2$, with $X_i=\frac{7}{4}$, $X_{12}^{11}=X_{21}^{11}=-\frac{1}{5}$, $X_{21}^{12}=0$ and $\rho_i=20$. By Proposition~\ref{prop:an}, the interconnected system is well-posed, asymptotically stable and the expression $\|\mathcal{P}_\mathcal{I}\|_{\mathscr{H}_2}<\gamma$ holds for all $\gamma>\sqrt{X_1+X_2}=\sqrt{\frac{7}{2}}\approx 1.87$. The actual $\mathscr{H}_2$ norm of the system is $\|\mathcal{P}_\mathcal{I}\|_{\mathscr{H}_2}=1.68$.
\end{example}


\section{Distributed $\mathscr{H}_2$ controller synthesis}
Consider the case where each subsystem $\mathcal{P}_i$ has a control input $u_i$ and a measured output $y_i$, such that
\begin{align} \label{eq:subsysuy}
\begin{pmatrix}
x_i(k+1)\\ o_i(k)\\ z_i(k)\\ y_i(k)
\end{pmatrix}=\begin{pmatrix}
A_i^{\mathrm{T}\mathrm{T}} & A_i^{\mathrm{T}\mathrm{S}} & B_i^{\mathrm{T}d} & B_i^{\mathrm{T}u}\\ A_i^{\mathrm{S}\mathrm{T}} & A_i^{\mathrm{S}\mathrm{S}} & B_i^{\mathrm{S}d} & B_i^{\mathrm{S}u}\\ C_i^{z\mathrm{T}} & C_i^{z\mathrm{S}} & D_i^{zd} & D_i^{zu}\\ C_i^{y\mathrm{T}} & C_i^{y\mathrm{S}} & D_i^{yd} & D_i^{yu}
\end{pmatrix}\begin{pmatrix}
x_i(k)\\ s_i(k)\\ d_i(k)\\ u_i(k)
\end{pmatrix},
\end{align}
where we assume that $D_i^{yu}=0$, without loss of generality \citep{langbortetal2004}.

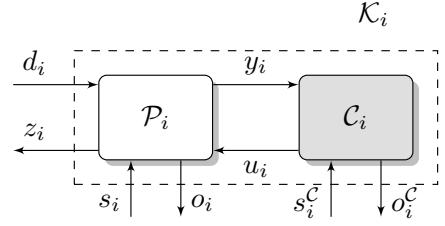
\begin{figure}[!t]
\centering
{\resizebox{2.4in}{!}{
\begin{tikzpicture}[auto,>=latex',node distance = 1.5em]
\matrix[ampersand replacement= \|, row sep=3em, column sep =1.5em](M1) {
\node [var,draw = none](w1) {};
\|
\node [block, drop shadow](P1) {$\mathcal{P}_i$};
\|\|
\node [cblock, drop shadow](C1) {$\mathcal{C}_i$};
\\
};

\node [node distance = 3em, left = of P1.150] (w11){}; \node [node distance = 3em, left = of P1.210] (w13){};

\node [right = of P1.30] (c11){}; \node [right = of P1.0] (c12){}; \node [right = of P1.330] (c13){};

\node [node distance = 2em, below = of P1.240] (s1){}; \node [node distance = 2em, below = of P1.300] (o1){};

\draw[->] (s1) -- node[near start]{$s_i$}(P1.240); \draw[->]  (P1.300) -- node[near end]{$o_i$}(o1);

\draw[->] (w11) -- node[near start]{$d_i$}(P1.150); \draw[<-]  (w13)-- node[near start]{$z_i$}(P1.210);

%

%
%

\node [node distance = 2em, below = of C1.240] (s1c){}; \node [node distance = 2em, below = of C1.300] (o1c){};

\draw[->] (s1c) -- node[near start]{$s_i^\mathcal{C}$}(C1.240); \draw[->]  (C1.300) -- node[near end]{$o_i^\mathcal{C}$}(o1c);

\draw[->] (P1.30) -- node[]{$y_i$} (C1.150); \draw[->] (C1.210) -- node[]{$u_i$}(P1.330);

\node[outerblock, fill opacity =0, fit=(P1) (C1)] (K1) {};
\node [above =.5em of K1.30] {$\mathcal{K}_i$};
\end{tikzpicture}
}}
\caption{Interconnection visualization of a locally controlled system $\mathcal{K}_i$, $i\in\mathbb{Z}_{[1:L]}$.}
\label{fig:Ki}
\end{figure}

The to-be-synthesized distributed controller is also an interconnected system, with subsystems $\mathcal{C}_i$, $i\in\mathbb{Z}_{[1:L]}$, described by
\begin{align}
\begin{pmatrix}
\xi_i(k+1)\\ o_i^\mathcal{C}(k) \\ u_i(k)
\end{pmatrix}=\begin{pmatrix}
(A_i^{\mathrm{T}\mathrm{T}})_\mathcal{C} & (A_i^{\mathrm{T}\mathrm{S}})_\mathcal{C}  & (B_i^\mathrm{T})_\mathcal{C} \\ (A_i^{\mathrm{S}\mathrm{T}})_\mathcal{C}  & (A_i^{\mathrm{S}\mathrm{S}})_\mathcal{C}  & (B_i^\mathrm{S})_\mathcal{C} \\ (C_i^\mathrm{T})_\mathcal{C}  & (C_i^\mathrm{S})_\mathcal{C}  & (D_i)_\mathcal{C} 
\end{pmatrix}\begin{pmatrix}
\xi_i(k)\\ s_i^\mathcal{C}(k)\\ y_i(k)
\end{pmatrix},
\end{align}
where $\xi_i:\mathbb{Z}\to \mathbb{R}^{k_{i}}$ is the controller's state, and $o_i^\mathcal{C}:\mathbb{Z}\to \mathbb{R}^{n_i^\mathcal{C}}$, $s_i^\mathcal{C}:\mathbb{Z}\to\mathbb{R}^{n_i^\mathcal{C}}$ are the controller's interconnection (communication) variables. Controller $\mathcal{C}_i$ and $\mathcal{C}_j$ are interconnected only if $\mathcal{P}_i$ and $\mathcal{P}_j$ are interconnected and the interconnection equation is
\begin{align} \label{eq:intC}
\begin{pmatrix}
o_{ij}^\mathcal{C}(k)\\s_{ij}^\mathcal{C}(k)
\end{pmatrix}=\begin{pmatrix}
s_{ji}^\mathcal{C}(k)\\o_{ji}^\mathcal{C}(k)
\end{pmatrix},\quad \forall k\in\mathbb{Z}.
\end{align}
The local closed-loop (controlled) system, $\mathcal{K}_i$ say, can then be represented by
\begin{align} \label{eq:subsysK}
\begin{pmatrix}
x_i^\mathcal{K}(k+1)\\o_i^\mathcal{K}(k)\\ z_i(k)
\end{pmatrix}=\underbrace{\begin{pmatrix}
(A_i^{\mathrm{T}\mathrm{T}})_\mathcal{K} & (A_i^{\mathrm{T}\mathrm{S}})_\mathcal{K}  & (B_i^\mathrm{T})_\mathcal{K} \\ (A_i^{\mathrm{S}\mathrm{T}})_\mathcal{K}  & (A_i^{\mathrm{S}\mathrm{S}})_\mathcal{K}  & (B_i^\mathrm{S})_\mathcal{K} \\ (C_i^\mathrm{T})_\mathcal{K}  & (C_i^\mathrm{S})_\mathcal{K}  & (D_i)_\mathcal{K} 
\end{pmatrix}}_{=:\Gamma_i}\begin{pmatrix}
x_i^\mathcal{K}(k)\\ s_i^\mathcal{K}(k)\\d_i(k)
\end{pmatrix},
\end{align}
where $x_i^\mathcal{K}:=\operatorname{col}(x_i,\xi_i)$, $o_i^\mathcal{K}:=\operatorname{col}(o_i,o_i^\mathcal{C})$ and $s_i^\mathcal{K}:=\operatorname{col}(s_i,s_i^\mathcal{C})$. Such a representation is obtained through elimination of the control variables $y_i$, $u_i$, as depicted in Figure~\ref{fig:Ki}. The state-space matrices of a closed-loop subsystem are affine with respect to the state-space matrices of the local controller:
\begin{align} \label{eq:Gami}
\Gamma_i=U_i^\top \Theta_i V_i+W_i,
\end{align}
with
\small
\begin{align*}
\Theta_i&:=\begin{pmatrix}
(A_i^{\mathrm{T}\mathrm{T}})_\mathcal{C} & (A_i^{\mathrm{T}\mathrm{S}})_\mathcal{C}  & (B_i^\mathrm{T})_\mathcal{C} \\ (A_i^{\mathrm{S}\mathrm{T}})_\mathcal{C}  & (A_i^{\mathrm{S}\mathrm{S}})_\mathcal{C}  & (B_i^\mathrm{S})_\mathcal{C} \\ (C_i^\mathrm{T})_\mathcal{C}  & (C_i^\mathrm{S})_\mathcal{C}  & (D_i)_\mathcal{C} 
\end{pmatrix}\!,\ \! V_i:=\begin{pmatrix}
0 & I & 0 & 0 & 0\\ 0 & 0 & 0 & I & 0\\ C_i^{y\mathrm{T}} & 0 & C_i^{y\mathrm{S}}  & 0 & D_i^{yd}
\end{pmatrix}\!,\\
U_i^\top&:=\begin{pmatrix}
0 & 0 & B_i^{\mathrm{T}u}\\ I & 0 & 0\\ 0 & 0 & B_i^{\mathrm{S}u}\\ 0 & I & 0\\ 0 & 0 & D_i^{zu}
\end{pmatrix}\!,\ W_i:=\begin{pmatrix}
A_i^{\mathrm{T}\mathrm{T}} & 0 & A_i^{\mathrm{T}\mathrm{S}} & 0 & B_i^{\mathrm{T}d}\\ 0 & 0 & 0 & 0 & 0\\ A_i^{\mathrm{S}\mathrm{T}} & 0 & A_i^{\mathrm{S}\mathrm{S}} & 0 & B_i^{\mathrm{S}d}\\ 0 & 0 & 0 & 0 & 0\\ C_i^{z\mathrm{T}} & 0 & C_i^{z\mathrm{S}} & 0 & D_i^{zd}
\end{pmatrix}\!.
\end{align*}
\normalsize

The feasibility test provided by Proposition \ref{prop:an} directly induces a feasibility test for well-posedness, stability and $\mathscr{H}_2$ performance for the closed-loop system, which consists of subsystems \eqref{eq:subsysK}, as stated in the following corollary. Define the matrix
\begin{align*}
T_i^\mathcal{K}:=\left(\begin{array}{ccc}
I & 0 & 0\\ (A_i^{\mathrm{T}\mathrm{T}})_\mathcal{K} & (A_i^{\mathrm{T}\mathrm{S}})_\mathcal{K}  & (B_i^\mathrm{T})_\mathcal{K} \\ \hline
(A_i^{\mathrm{S}\mathrm{T}})_\mathcal{K}  & (A_i^{\mathrm{S}\mathrm{S}})_\mathcal{K}  & (B_i^\mathrm{S})_\mathcal{K} \\ 0 & I & 0\\ \hline
(C_i^\mathrm{T})_\mathcal{K}  & (C_i^\mathrm{S})_\mathcal{K}  & (D_i)_\mathcal{K} \\
0 & 0 & I
\end{array}\right).
\end{align*}

\begin{corollary} \label{cor:cl}
The interconnected system $\mathcal{K}_\mathcal{I}$ of \eqref{eq:subsysK} is well-posed, AS and $\|\mathcal{K}_\mathcal{I}\|_{\mathscr{H}_2}<\gamma$, if $(B_i^\mathrm{S})_\mathcal{K}=0$ for all $i\in\mathbb{Z}_{[1:L]}$ and there exist positive-definite $X_i^\mathcal{K}\in\mathbb{R}^{2k_{i}\times 2k_{i}}$, $\rho_i>0$, symmetric $(X_{ij}^{11})_\mathcal{K}\in\mathbb{R}^{(n_{ij}+n_{ij}^\mathcal{C})\times(n_{ij}+n_{ij}^\mathcal{C})}$, $(i,j)\in\mathbb{Z}_{[1:L]}^2$, and $(X_{ij}^{12})_\mathcal{K}\in\mathbb{R}^{(n_{ij}+n_{ij}^\mathcal{C})\times(n_{ij}+n_{ij}^\mathcal{C})}$, $(i,j)\in\mathbb{Z}_{[1:L]}^2$, $i>j$, with
\begin{align}
\arraycolsep=2pt
&(T_i^\mathcal{K})^\top\left(\begin{array}{cc|cc|cc}
\!\!-X_i^\mathcal{K} \!\!&\!\! 0 & 0 \!\!\!&\!\!\! 0 & 0 & 0\\
0 \!\!&\!\! X_i^\mathcal{K} & 0 \!\!\!&\!\!\! 0& 0 & 0\\ \hline
0 \!\!&\!\! 0 & (Z_i^{11})_\mathcal{K} \!&\! (Z_i^{12})_\mathcal{K} & 0 & 0\\ 0 \!\!&\!\! 0 & (Z_i^{12})_\mathcal{K}^\top \!&\! (Z_i^{22})_\mathcal{K} & 0& 0\\ \hline
0 \!\!&\!\! 0 & 0 \!\!&\!\! 0 & I & 0\\ 0 \!\!&\!\! 0 & 0 \!\!&\!\! 0 & 0 & -\rho_i I
\end{array}\right)T_i^\mathcal{K}\prec 0,\label{eq:clineq}\\
& \sum_{i=1}^L \operatorname{trace} \left( (B_i^\mathrm{T})_\mathcal{K}^\top X_i^\mathcal{K} (B_i^\mathrm{T})_\mathcal{K}+(D_i)_\mathcal{K}^\top (D_i)_\mathcal{K}\right)<\gamma^2,\label{eq:clperf}
\end{align}
where
\begin{align*}
(Z_i^{11})_\mathcal{K}&:=\begin{pmatrix}
(Z_i^{11})_\mathcal{P} & (Z_i^{11})_{\mathcal{PC}}\\ (Z_i^{11})^\top_{\mathcal{PC}} & (Z_i^{11})_\mathcal{C}
\end{pmatrix},\\
(Z_i^{12})_\mathcal{K}&:=\begin{pmatrix}
(Z_i^{12})_\mathcal{P} & (Z_i^{12})_{\mathcal{PC}}\\ (Z_i^{12})_{\mathcal{CP}} & (Z_i^{12})_\mathcal{C}
\end{pmatrix},\\
(Z_i^{22})_\mathcal{K}&:=\begin{pmatrix}
(Z_i^{22})_\mathcal{P} & (Z_i^{22})_{\mathcal{PC}}\\ (Z_i^{22})^\top_{\mathcal{PC}} & (Z_i^{22})_\mathcal{C}
\end{pmatrix},
\end{align*}
with the submatrices defined in Appendix \ref{app:clscales}.
\end{corollary}

Recall the definition of $T_i$ in \eqref{eq:Ti} and define
\begin{align*}
S_i= \left(\begin{array}{ccc}
(A_i^{\mathrm{TT}})^\top &(A_i^{\mathrm{ST}})^\top & (C_i^{z\mathrm{T}})^\top\\
- I & 0 & 0\\ \hline
0 & -I & 0\\
(A_i^{\mathrm{TS}})^\top & (A_i^{\mathrm{SS}})^\top & (B_i^{\mathrm{S}d})^\top\\ \hline
0 & 0 & -I\\
(B_i^{\mathrm{T}d})^\top & (B_i^{\mathrm{S}d})^\top & (D_i^{zd})^\top
\end{array}\right).
\end{align*}

\subsection{Convex distributed $\mathscr{H}_2$ controller existence conditions}\
We are now ready to state the main result, which provides necessary and sufficient conditions for the existence of a distributed controller that satisfies the conditions in Corollary~\ref{cor:cl}, in the form of LMIs.

\begin{proposition} \label{prop:exist}
Let $B_i^{\mathrm{S}d}=0$, $D_i^{yd}=0$ for all $i\in\mathbb{Z}_{[1:L]}$. The following statements are equivalent:
\begin{itemize}
\item There exist controllers $\mathcal{C}_i$, with $n_{ij}^\mathcal{C}=3n_{ij}$ for all $(i,j)\in\mathbb{Z}_{[1:L]}^2$ so that the controlled interconnected system described by  \eqref{eq:int}, \eqref{eq:intC} and \eqref{eq:subsysK} admits $\rho_i>0$, matrices $X_i^\mathcal{K}\succ 0$, $i\in\mathbb{Z}_{[1:L]}$, symmetric $(X_{ij}^{11})_\mathcal{K}$, $(i,j)\in\mathbb{Z}_{[1:L]}^2$, and $(X_{ij}^{12})_\mathcal{K}$, $(i,j)\in\mathbb{Z}_{[1:L]}^2$, $i>j$, that satisfy inequalities \eqref{eq:clineq} and \eqref{eq:clperf}.
\vspace{1em}
\item There exist $X_i$, $Y_i$, symmetric $(X_{ij}^{11})_\mathcal{P}$, $(Y_{ij}^{11})_\mathcal{P}$, $\alpha_i,\beta_i>0$ for all $(i,j)\in\mathbb{Z}_{[1:L]}^2$, and $(X_{ij}^{12})_\mathcal{P}$, $(Y_{ij}^{12})_\mathcal{P}$ for all $(i,j)\in\mathbb{Z}_{[1:L]}^2$, $i>j$, that satisfy
\begin{align}
&\begin{pmatrix}
X_i & I\\ I & Y_i
\end{pmatrix}\succ 0,\label{eq:existXY}\\
&\sum_{i=1}^L \operatorname{trace} \left((B_i^{\mathrm{T}d})^\top X_i B_i^{\mathrm{T}d}+(D_i^{zd})^\top D_i^{zd}\right)<\gamma^2,\label{eq:existperf}
\end{align}
\begin{align}
&\Psi_i^\top T_i^\top\left(\begin{array}{cc|cc|cc}
\!\! -X_i & 0 & 0 & 0 & 0 & 0\!\!\\
\!\! 0 & X_i & 0 & 0& 0 & 0\!\!\\ \hline
\!\! 0 & 0 & (Z_i^{11})_\mathcal{P} & (Z_i^{12})_\mathcal{P} & 0 & 0\!\!\\
\!\! 0 & 0 & (Z_i^{12})^\top_\mathcal{P} & (Z_i^{22})_\mathcal{P} & 0 & 0
\!\!\\ \hline
\!\! 0 & 0 & 0 & 0 & I & 0\!\!\\
\!\! 0 & 0 & 0 & 0 & 0 & -\alpha_i I\!\!
\end{array}\right)T_i\Psi_i\prec 0, \label{eq:existX}
\end{align}
\begin{align}
&\Phi_i^\top S_i^\top \left(\begin{array}{cc|cc|cc}
\!\! -Y_i & 0 & 0 & 0 & 0 & 0\!\!\\
\!\! 0 & Y_i & 0 & 0& 0 & 0\!\!\\ \hline
\!\! 0 & 0 & (W_i^{11})_\mathcal{P} & (W_i^{12})_\mathcal{P} & 0 & 0\!\!\\
\!\! 0 & 0 & (W_i^{12})^\top_\mathcal{P} & (W_i^{22})_\mathcal{P} & 0& 0\!\!\\ \hline
\!\! 0 & 0 & 0 & 0 & I & 0\!\!\\
\!\! 0 & 0 & 0 & 0 & 0 & -\beta_i I\!\!
\end{array}\right)S_i\Phi_i\succ 0,\label{eq:existY}
\end{align}
where the columns of $\Psi_i$ and $\Phi_i$ form a basis of $\operatorname{ker} (C_i^{yT}\ C_i^{yS}\ D_i^{yd})$ and $\operatorname{ker} ((B_i^{\mathrm{T}u})^\top\ (B_i^{\mathrm{S}u})^\top\  (D_i^{zu})^\top)$, respectively, and
\begin{align*}
(W_i^{11})_\mathcal{P} &:=-\!\!\operatorname*{diag}_{j\in\mathbb{Z}_{[1:L]}} (Y_{ij}^{11})_\mathcal{P},\, (W_i^{22})_\mathcal{P} :=\!\!\operatorname*{diag}_{j\in\mathbb{Z}_{[1:L]}} (Y_{ji}^{11})_\mathcal{P},\\
(W_i^{12})_\mathcal{P} &:=\operatorname{diag}\left(-\operatorname*{diag}_{j\in\mathbb{Z}_{[1:i]}} (Y_{ij}^{12})_\mathcal{P},\operatorname*{diag}_{j\in\mathbb{Z}_{[i+1,L]}} (Y_{ji}^{12})^\top_\mathcal{P}\right).
\end{align*}
\end{itemize}
\end{proposition}
\begin{pf}
We first show that the existence of positive scalars $\alpha_i$ and $\beta_i$ such that \eqref{eq:existX} and \eqref{eq:existY} hold is equivalent with the existence of a positive scalar $\rho_i$ such that
\begin{align} \label{eq:star}
\Psi_i^\top T_i^\top \Lambda_i(\rho_i)T_i\Psi_i\prec 0 \text{ and } \Phi_i^\top S_i^\top \Pi_i(\rho_i^{-1})S_i\Phi_i\succ 0,
\end{align}
with
\begin{align*}
&\Lambda_i:\xi\mapsto \operatorname{diag}(-X_i,X_i,(Z_i)_\mathcal{P},I,-\xi I) \text{ and }\\
&\Pi_i: \xi \mapsto \operatorname{diag}(-Y_i,Y_i,(W_i)_\mathcal{P},I,-\xi I).
\end{align*}
For sufficiency, let $\alpha_i$ and $\beta_i$ satisfy \eqref{eq:existX} and \eqref{eq:existY}. We distinguish two cases. First, if $\alpha_i\beta_i\geq 1$, then
\begin{align*}
&\underbrace{\Phi_i^\top S_i^\top \Pi_i(\beta_i)S_i\Phi_i}_{\succ 0}+\underbrace{\Phi_i^\top S_i^\top \operatorname{diag}(0,0,0,0,(\beta_i-\alpha_i^{-1})I)S_i\Phi_i}_{\succeq 0}\\
&=\Phi_i^\top S_i^\top \Pi_i(\alpha_i^{-1})S_i\Phi_i\succ 0.
\end{align*}
Hence, \eqref{eq:star} holds for $\rho_i=\alpha_i$. In the other case $\alpha_i\beta_i<1$, thus it follows that
\begin{align*}
&\underbrace{\Psi_i^\top T_i^\top \Lambda_i(\alpha_i)T_i\Psi_i}_{\prec 0} + \underbrace{\Psi_i^\top T_i^\top \operatorname{diag}(0,0,0,0,(\alpha_i-\beta_i^{-1})T_i\Psi_i}_{\preceq 0}\\
&=\Psi_i^\top T_i^\top \Lambda_i(\beta_i^{-1})T_i\Psi_i\prec 0.
\end{align*}
Hence, \eqref{eq:star} holds for $\rho_i=\beta_i^{-1}$. Necessity follows directly by taking $\alpha_i=\rho_i$ and $\beta_i=\rho_i^{-1}$.

For a proof that the existence of $X_i$, $Y_i$, $(Z_i)_\mathcal{P}$, $(W_i)_\mathcal{P}$ and $\rho_i$ that satisfy \eqref{eq:star} and \eqref{eq:existXY} is equivalent with the existence of $X_i^\mathcal{K}$, $(Z_i)_\mathcal{K}$ and $\rho_i$ that satisfy \eqref{eq:clineq}, we refer the reader to \citep{langbortetal2004} due to space limitations.

Finally, we will show that \eqref{eq:existperf} is equivalent with \eqref{eq:clperf}. We note that for necessity $X_i$ can be taken as the upper-left block of $X_i^\mathcal{K}$, while for sufficiency, $X_i^\mathcal{K}$ can be taken such that its upper-left block equals $X_i$ \citep{langbortetal2004}. Thus, by \eqref{eq:Gami}, we have that
\begin{align*}
(B_i^{\mathrm{T}d})^\top X_i B_i^{\mathrm{T}d}+(D_i^{zd})^\top D_i^{zd}&=(B_i^\mathrm{T})_\mathcal{K}^\top X_i^\mathcal{K} (B_i^\mathrm{T})_\mathcal{K}\\
&\quad+(D_i)_\mathcal{K}^\top (D_i)_\mathcal{K}
\end{align*}
for all $i\in\mathbb{Z}_{[1:L]}$, since $D_i^{yd}=0$. It therefore follows that \eqref{eq:existperf} $\Leftrightarrow$ \eqref{eq:clperf}, which concludes the proof.\hfill $\blacksquare$
\end{pf}

\begin{remark}
The equivalence between the convex conditions \eqref{eq:existX}, \eqref{eq:existY} and non-convex conditions \eqref{eq:star} can be transferred to the continuous-time case \citep[Theorem~2]{chen2019} \emph{mutatis mutandis}. The continuous-time distributed $\mathscr{H}_2$ controller existence problem can then be solved via equivalent LMIs, instead of the equivalent bilinear optimization problem with $L$ additional LMIs in \citep{chen2019}, with $L$ the cardinality of the vertex set $V$.
\end{remark}

\subsection{Decentralized $\mathscr{H}_2$ controller existence conditions}
\begin{figure}[!t]
\centering
{\resizebox{2.4in}{!}{
\begin{tikzpicture}[auto,>=latex',node distance = 1.5em]
\matrix[ampersand replacement= \|, row sep=3em, column sep =1.5em](M1) {
\node [var,draw = none](w1) {};
\|
\node [block, drop shadow](P1) {$\mathcal{P}_i$};
\|\|
\node [cblock, drop shadow](C1) {$\mathcal{C}_i$};
\\
};

\node [node distance = 3em, left = of P1.150] (w11){}; \node [node distance = 3em, left = of P1.210] (w13){};

\node [right = of P1.30] (c11){}; \node [right = of P1.0] (c12){}; \node [right = of P1.330] (c13){};

\node [node distance = 2em, below = of P1.240] (s1){}; \node [node distance = 2em, below = of P1.300] (o1){};

\draw[->] (s1) -- node[near start]{$s_i$}(P1.240); \draw[->]  (P1.300) -- node[near end]{$o_i$}(o1);

\draw[->] (w11) -- node[near start]{$d_i$}(P1.150); \draw[<-]  (w13)-- node[near start]{$z_i$}(P1.210);

%

%
%

%

\draw[->] (P1.30) -- node[]{$y_i$} (C1.150); \draw[->] (C1.210) -- node[]{$u_i$}(P1.330);

\node[outerblock, fill opacity =0, fit=(P1) (C1)] (K1) {};
\node [above =.5em of K1.30] {$\mathcal{K}_i$};
\end{tikzpicture}
}}
\caption{Locally controlled system $\mathcal{K}_i$ for a decentralized controller.}
\label{fig:Kidec}
\end{figure}
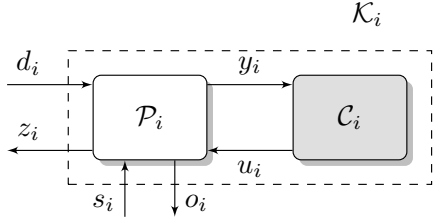

A special distributed controller is a decentralized controller, where no controller interconnections are present. This is depicted in Figure~\ref{fig:Kidec} for a locally controlled system. The synthesis of decentralized controllers is motivated by interconnected systems where no communication between controllers is possible. In this case $n_i^\mathcal{C}=0$, hence Proposition~\ref{prop:exist} cannot be applied for the construction of a decentralized controller, since it guarantees the existence of a controller with $n_{ij}^\mathcal{C}=3n_{ij}$ only.

Therefore, we provide conditions for the existence of a controller with $n_{ij}^\mathcal{C}=0$ which achieves global $\mathscr{H}_2$ performance by fixing the supply functions related to the interconnection variables. Given symmetric $X_{ij}^{11}$, $(i,j)\in\mathbb{Z}_{[1:L]}^2$, and $X_{ij}^{12}$, $(i,j)\in\mathbb{Z}_{[1:L]}^2$, $i>j$, we have the following result.

\begin{proposition} \label{prop:existdec}
Let $B_i^{\mathrm{S}d}=0$, $D_i^{yd}=0$ for all $i\in\mathbb{Z}_{[1:L]}$. The following statements are equivalent:
\begin{itemize}
\item There exist controllers $\mathcal{C}_i$, with $n_{i}^\mathcal{C}=0$ for all $i\in\mathbb{Z}_{[1:L]}$ so that the controlled interconnected system described by \eqref{eq:int}, \eqref{eq:intC} and \eqref{eq:subsysK} admits $\rho_i\in\mathbb{R}_{>0}$, $X_i^\mathcal{K}\succ 0$, $i\in\mathbb{Z}_{[1:L]}$ that satisfy \eqref{eq:clineq} and \eqref{eq:clperf} for $(X_{ij}^{11})_\mathcal{K}=X_{ij}$ and $(X_{ij}^{12})_\mathcal{K}=X_{ij}^{12}$.
\vspace{1em}
\item There exist $X_i$, $Y_i$ and $\alpha_i,\beta_i>0$, $i\in\mathbb{Z}_{[1:L]}$, so that \eqref{eq:existXY}, \eqref{eq:existperf}, \eqref{eq:existX} and \eqref{eq:existY} are satisfied for $(Z_i^{11})_\mathcal{P}=Z_i^{11}$, $(Z_i^{12})_\mathcal{P}=Z_i^{12}$, $(Z_i^{22})_\mathcal{P}=Z_i^{22}$ and 
\begin{align} \label{eq:decinv}
\begin{pmatrix}
(W_i^{11})_\mathcal{P} & (W_i^{12})_\mathcal{P}\\ (W_i^{12})^\top_\mathcal{P} & (W_i^{22})_\mathcal{P}
\end{pmatrix}:=\begin{pmatrix}
Z_i^{11} & Z_i^{12}\\ (Z_i^{12})^\top & Z_i^{22}
\end{pmatrix}^{-1}.
\end{align}
\end{itemize}
\end{proposition}
\begin{pf}
($\Leftarrow$) Take an arbitrary $i\in\mathbb{Z}_{[1:L]}$. By \eqref{eq:existXY}, there exist extended matrices $X_i^\mathcal{K}$, $Y_i^\mathcal{K}$, so that $X_i^\mathcal{K}=(Y_i^\mathcal{K})^{-1}$. Define $\Lambda_i:=\operatorname{diag}(-X_i^\mathcal{K},X_i^\mathcal{K},Z_i,I,-\rho_iI)$. Then by \eqref{eq:existX} and \eqref{eq:existY}, a permutation of $\Lambda_i$ gives a matrix $P_i$ which satisfies
\begin{align} \label{eq:elim2}
(V_i)^\top_\bot \begin{pmatrix}
I\\ W_i
\end{pmatrix}^\top P_i\begin{pmatrix}
I\\ W_i
\end{pmatrix}(V_i)_\bot&\prec 0\text{ and } \nonumber\\
(U_i)^\top_\bot\begin{pmatrix}
-W_i^\top\\ I
\end{pmatrix}^\top P_i^{-1}\begin{pmatrix}
-W_i^\top\\ I
\end{pmatrix}(U_i)_\bot&\succ 0.
\end{align}
Hence, by the elimination lemma \citep{scherer2001}, there exists a $\Theta_i$ so that
\begin{align} \label{eq:elim1}
\begin{pmatrix}
I\\ U_i^\top \Theta_iV_i+W_i
\end{pmatrix}^\top P_i\begin{pmatrix}
I\\ U_i^\top \Theta_iV_i+W_i
\end{pmatrix}\prec 0,
\end{align}
which is equivalent with \eqref{eq:clineq} for $(X_{ij}^{11})_\mathcal{K}=X_{ij}$ and $(X_{ij}^{12})_\mathcal{K}=X_{ij}^{12}$.

($\Rightarrow$) To show necessity, observe again that \eqref{eq:clineq} is equivalent with \eqref{eq:elim1}, which is equivalent with \eqref{eq:elim2}. Then, by taking $X_i$ and $Y_i$ as the upper-left blocks of $X_i^\mathcal{K}$ and $Y_i^\mathcal{K}$, respectively, we obtain \eqref{eq:existX} and \eqref{eq:existY}.

The equivalence of \eqref{eq:existperf} and \eqref{eq:clperf} was shown in the proof of Proposition~\ref{prop:exist}, which concludes the proof.\hfill $\blacksquare$
\end{pf}

\begin{remark}
Fixing the supply functions for the closed-loop subsystems as $\sigma_{ij}(s_{ij},o_{ij})=o_{ij}^\top s_{ij}$, corresponding to $X_{ij}^{11}=0$ and $X_{ij}^{12}=\frac{1}{2}I$, implies that the closed-loop subsystems are required to be passive with respect to the interconnection variables. The design of passive systems holds an important place in control theory \citep{vanderschaft2016} and is a classical method for guaranteeing stability of interconnected systems \citep{arcak2016}; see e.g. \citep{cucuzzella2019} for a recent development of passivity-based distributed control for DC microgrids.
\end{remark}

\subsection{Controller construction}
\begin{algorithmIF} \label{alg}
For each pair $(i,j)\in\mathbb{Z}_{[1:L]}^2$, let $X_i$, $Y_i$, $\rho_i$, $(X_{ij}^{11})_\mathcal{P}$, $(Y_{ij}^{11})_\mathcal{P}$, and for each pair $(i,j)\in\mathbb{Z}_{[1:L]}^2$, $i>j$, let $(X_{ij}^{12})_\mathcal{P}$, $(Y_{ij}^{12})_\mathcal{P}$, be computed to satisfy \eqref{eq:existXY}, \eqref{eq:existperf}, \eqref{eq:existX}, \eqref{eq:existY}. For decentralized control, let \eqref{eq:decinv} be satisfied, additionally.

 For each $i\in\mathbb{Z}_{[1:L]}$, the synthesis of controller $\mathcal{C}_i$ proceeds as follows (skip step (2) for decentralized control): 
\begin{enumerate}
\item Let $M_i$ and $N_i$ be non-singular and such that $M_iN_i^\top=I-X_iY_i$. Compute $X_i^\mathcal{K}$ as the unique solution to the linear equation
\begin{align*}
X_i^\mathcal{K}\begin{pmatrix}
Y_i & I\\ N_i^\top & 0
\end{pmatrix}=\begin{pmatrix}
I & X_i\\ 0 & M_i^\top
\end{pmatrix}.
\end{align*}
\item Define 
\small
\begin{align*}
X_{ij}^\mathcal{P}&:=\begin{pmatrix}
(X_{ij}^{11})_\mathcal{P} & (X_{ij}^{12})_\mathcal{P}\\ (X_{ij}^{12})_\mathcal{P}^\top & -(X_{ji}^{11})_\mathcal{P}
\end{pmatrix},\,Y_{ij}^\mathcal{P}:=\begin{pmatrix}
(Y_{ij}^{11})_\mathcal{P} & (Y_{ij}^{12})_\mathcal{P}\\ (Y_{ij}^{12})_\mathcal{P}^\top & -(Y_{ji}^{11})_\mathcal{P}
\end{pmatrix}.
\end{align*}
\normalsize
and compute an eigendecomposition $X_{ij}^\mathcal{P}-(Y_{ij}^\mathcal{P})^{-1}=V_{ij}\Lambda_{ij}V_{ij}^\top$, with $\Lambda_{ij}$ a diagonal matrix with the eigenvalues on its diagonal in a descending order. Scale the eigenvectors as $\bar{V}_{ij}=V_{ij}|\Lambda_{ij}|^{\frac{1}{2}}$ such that 
\begin{align*}
X_{ij}^\mathcal{P}-(Y_{ij}^\mathcal{P})^{-1}=(\bar{V}_{ij}^{+}\ \bar{V}_{ij}^{-})\operatorname{diag}(I,-I)(\bar{V}_{ij}^{+}\ \bar{V}_{ij}^{-})^\top,
\end{align*}
with $\bar{V}_{ij}=:(\bar{V}_{ij}^{+}\ \bar{V}_{ij}^{-})$. Let $M_{ij}^{22}:=\operatorname{diag}(I_{3n_{ij}},-I_{3n_{ij}})$ and $M_{ij}^{12}:=\frac{1}{\sqrt{3}}(
\bar{V}_{ij}^{+}\, \bar{V}_{ij}^{+}\,  \bar{V}_{ij}^{+}\, \bar{V}_{ij}^{-}\, \bar{V}_{ij}^{-}\, \bar{V}_{ij}^{-})$ and define
\begin{align*}
M_{ij}^{12}&=:\begin{pmatrix}
(X_{ij}^{11})_\mathcal{PC} & (X_{ij}^{12})_\mathcal{PC}\\ (X_{ij}^{12})_\mathcal{CP}^\top & -(X_{ji}^{11})_\mathcal{PC}
\end{pmatrix},\\
M_{ij}^{22}&=:\begin{pmatrix}
(X_{ij}^{11})_\mathcal{C} & (X_{ij}^{12})_\mathcal{C}\\ (X_{ij}^{12})_\mathcal{C}^\top & -(X_{ji}^{11})_\mathcal{C}
\end{pmatrix}.
\end{align*}
\item Construct the closed-loop scales defined in Appendix~\ref{app:clscales} and let
\begin{align*}
P_i:=\left(\begin{array}{ccc|ccc}
-X_i^\mathcal{K} & 0 & 0 & 0 & 0 & 0\\ 0 & (Z_i^{22})_\mathcal{K} & 0 & 0 & (Z_i^{12})_\mathcal{K}^\top & 0\\0 & 0 & -\rho_iI & 0& 0 & 0\\ \hline 0 & 0 & 0 & X_i^\mathcal{K} & 0 & 0\\ 0 & (Z_i^{12})_\mathcal{K} & 0 & 0 & (Z_i^{11})_\mathcal{K} & 0\\ 0 & 0 & 0 & 0 & 0 & I
\end{array}\right).
\end{align*}
Solve the following inequality for $\Theta_i$:
\begin{align} \label{eq:thetain}
\begin{pmatrix}
I\\ U_i^\top \Theta_iV_i+W_i
\end{pmatrix}^\top P_i\begin{pmatrix}
I\\ U_i^\top \Theta_iV_i+W_i
\end{pmatrix}\prec 0.
\end{align}
The quadratic matrix inequality \eqref{eq:thetain} can be solved by computing an eigendecomposition and a linear equation, see e.g. \citep{scherer2001} for details.
\end{enumerate}
\end{algorithmIF}

\section{Numerical examples}
To illustrate the distributed $\mathscr{H}_2$ controller synthesis method, we consider a linear coupled-oscillator network consisting of $L$ oscillators. For each node $i\in\mathbb{Z}_{[1:L]}$, the dynamics are described by
\begin{align} \label{eq: osc}
\quad m_i\ddot{\theta}_i+b_i\dot{\theta}_i=u_i-\sum_{j\in \mathcal{N}_i} k_{ij} (\theta_i-\theta_j)+d_i,
\end{align}
with inertia $m_i$, damping $b_i$ and coupling coefficient $k_{ij}=k_{ji}$. The mechanical analogue of a linear coupled-oscillator network is a network of masses that are interconnected through linear springs and have linear damping. A typical system that is modeled as a linear oscillator network is a linearized power network, consisting of generators ($m_i\neq 0$) and loads ($m_i=0$) \citep{bergen81, dorfler2012}. The local measurement is assumed to be $y_i:=\theta_i$ and the performance output is set equal to the state $z_i:=x_i:=\operatorname{col}(\theta_i,\dot{\theta}_i)$. We use a zero-order hold discretization with sampling time $T=0.1$ seconds for each subsystem and an approximation $e^{M}\approx I+M$, so that each subsystem $\mathcal{P}_i$ has an input/state/output representation~\eqref{eq:subsys} with matrices
\begin{align*}
A_i^\mathrm{TT}&=\begin{pmatrix}
1 & T\\ -\sum_{j\in\mathcal{N}_i} \frac{k_{ij}}{m_i}T & 1-\frac{b_i}{m_i}T
\end{pmatrix}, A_i^\mathrm{TS}=\operatorname*{row}_{j\in\mathcal{N}_i}\begin{pmatrix}
0 \\ \frac{k_{ij}}{m_i}T
\end{pmatrix},\\
A_i^\mathrm{ST}&=C_i^{y\mathrm{T}}=\operatorname*{col}_{j\in\mathcal{N}_i}\begin{pmatrix}
1 & 0
\end{pmatrix},\ A_i^\mathrm{SS}=0_{n_i\times n_i},\\ B_i^{\mathrm{S}d}&=B_i^{\mathrm{S}u}=0_{n_i\times 1},\ B_i^{\mathrm{T}d}=B_i^{\mathrm{T}u}=\operatorname{col}(0,\frac{T}{m_i}),\ C_i^{z\mathrm{T}}=I_2,\\
C_i^{z\mathrm{S}}&=0_{2\times n_i}, D_i^{zd}= D_i^{zu}=0_{2\times 1}, D_i^{yd} = D_i^{yu}=0.
\end{align*}

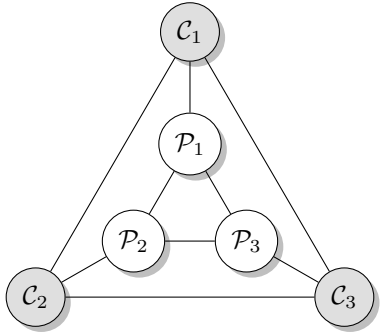
\begin{figure}[!t]
\centering
{\resizebox{2in}{!}{
\begin{tikzpicture}[auto,>=latex']
\node [circ, drop shadow](P2) {$\mathcal{P}_2$};
\node [circ, drop shadow](P1) at ($ (P2) + (60:1.5) $) {$\mathcal{P}_1$};
\node [circ, drop shadow](P3) at ($ (P2) + (0:1.5) $) {$\mathcal{P}_3$};
\node [ccirc,drop shadow](C1) at ($ (P1) + (90:1.5) $) {$\mathcal{C}_1$};
\node [ccirc,drop shadow](C2) at ($ (P2) + (210:1.5) $) {$\mathcal{C}_2$};
\node [ccirc,drop shadow](C3) at ($ (P3) + (330:1.5) $) {$\mathcal{C}_3$};
\draw[-] (P1) -- (P2); \draw[-] (P2) -- (P3); \draw[-] (P3) -- (P1);
\draw[-] (C1) -- (C2); \draw[-] (C2) -- (C3); \draw[-] (C3) -- (C1);
\draw[-] (P1) -- (C1); \draw[-] (P2) -- (C2); \draw[-] (P3) -- (C3);

\end{tikzpicture}
}}
\caption{Structure of the oscillator network represented by a triangle graph $(L=3$). The synthesized distributed $\mathscr{H}_2$ controller modules are depicted in gray.}
\label{fig:examplering}
\end{figure}

\begin{figure}[!t]
\centering
\hspace*{-1em}\includegraphics[width=3.75in]{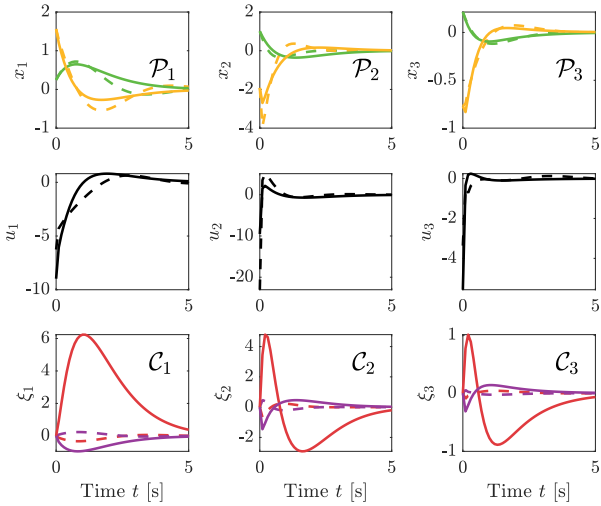}
\caption{Subsystem states $[x_i]_1$ (green) and $[x_i]_2$ (yellow), controller states $[\xi_i]_1$ (red) and $[\xi_i]_2$ (violet) and control inputs $u_i$ (black), $i\in\{1,2,3\}$, for the distributed (solid) and central (dashed) controller.}
\label{fig:triangle}
\end{figure}

\begin{figure}[!t]
\centering
\includegraphics[width=3.5in]{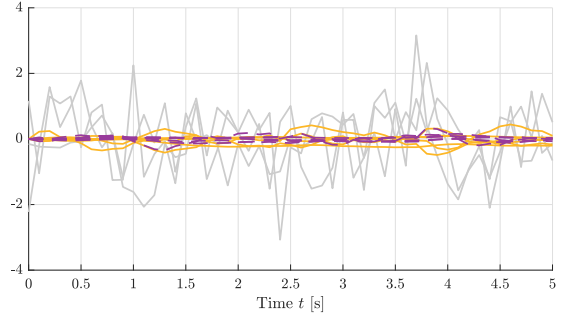}
\caption{Disturbances $d_i$ (gray) and corresponding performance output components $[z_i]_1$ and $[z_i]_2$, for the distributed (yellow) and centralized (purple) control.}
\label{fig:distrej}
\end{figure}

Let us consider a network with a triangular structure, as depicted in Figure \ref{fig:examplering}. The systems' inertia, damping and coupling coefficients are $m_1=3$, $m_2=1$, $m_3=2$, $b_1=2$, $b_2=1$, $b_3=4$ and $k_{12}=k_{23}=k_{31}=1$. The open-loop system is not AS. We aim for disturbance attenuation via the synthesis of a distributed controller that achieves unit $\mathscr{H}_2$ performance for the controlled network. We therefore verify the feasibility of the LMIs in Proposition~\ref{prop:exist} for $\gamma=1$. We find that the LMIs are feasible, hence there exists a distributed controller that achieves $\|\mathcal{K}_\mathcal{I}\|_{\mathscr{H}_2}<1$. The distributed controller is constructed according to Algorithm \ref{alg} and results in a closed-loop $\mathscr{H}_2$ norm of $0.22$. Due to space limitations, we refer the reader to \citep{steentjes2020} for a similar fully worked-out example, i.e.,  including numerical values for the controller matrices. Simulation of the controlled network with zero disturbance, with the subsystems' initial conditions drawn from a normal distribution $\mathcal{N}(0,1)$ and the controllers' initial conditions set identical to zero, results in the trajectories depicted in Figure~\ref{fig:triangle}. We observe that the subsystems' and controllers' states asymptotically converge to zero, illustrating asymptotic stability of the closed-loop system. For validation, we also compute a central controller via the feasibility problem in \citep{schererweilandLMI} for an $\mathscr{H}_2$ upper-bound equal to $0.22$. The resulting controller achieves an $\mathscr{H}_2$ norm of $0.18$ and the trajectories are shown in Figure~\ref{fig:triangle} (the central controller state $\xi\in\mathbb{R}^6$ is denoted $\xi=\operatorname{col}(\xi_1,\xi_2,\xi_3)$).

For illustration of the controlled network's ability to reduce output variance in the case of stochastic disturbance signals, we initialize the system with $x(0)=0$, $\xi(0)=0$, and apply signals $d_i$, that are mutually uncorrelated Gaussian white-noise processes with unit variance. Asymptotically, the obtained $\mathscr{H}_2$ norm for the controlled network is directly related to the output variance through $\lim_{k\to\infty} \mathrm{E} z^\top(k)z(k)=\|\mathcal{K}_\mathcal{I}\|_{\mathscr{H}_2}^2$ \citep{schererweilandLMI}. We therefore asses the variance of the output on a finite interval. Figure \ref{fig:distrej} shows the two components of the performance outputs $z_i$, which illustrate a significant attenuation of the stochastic disturbances by both the distributed and central controller.

\subsection{Computation times}
To demonstrate the scalability of the developed synthesis method, we consider the controller construction for the oscillator network on cycle graphs with increased values of $L$. For each graph, the constants $m_i$, $b_i$ and $k_{ij}=k_{ji}$ are drawn from uniform distributions $\mathcal{U}(1,2)$, $\mathcal{U}(2,3)$ and $\mathcal{U}(1,2)$, respectively. Table~\ref{table:comp} summarizes the times required to solve the controller existence LMIs in Proposition~\ref{prop:exist}. The performance bound is chosen as $\gamma=10$, such that the LMIs are feasible for all values of $L$ in Table~\ref{table:comp}. Computations were performed on a PC with Intel Core i5 at 2.3GHz and 16GB memory using MOSEK version 8.1. We observe that for a cycle graph of moderate size ($L=50$), the computation time is considerably lower for the distributed controller compared to the central controller. For $L\geq 100$, no solution was obtained for the central controller after 4 hours of computation, while the distributed controller problem was solved for up to $L=10,000$ in less than 6 seconds.

\begin{table}
\begin{center}
\caption{Computation times for solving the LMIs in Proposition~\ref{prop:exist} for the distributed $\mathscr{H}_2$ controller and the corresponding LMIs for the central $\mathscr{H}_2$ controller for a network of $L$ subsystems. $\dagger$: No solution after 4 hours.}
\begin{tabular}{||c|c|c||}
\hline
$L$ & Central controller & Distributed controller\\
\hline\hline
$3$ & 0.44s & 0.24s\\
$10$ & 0.78s & 0.29s\\
$50$ & 831.57s & 0.34s\\
$100$ & $\dagger$ & 0.42s\\
$1,000$ & $\dagger$ & 1.35s\\
$10,000$ & $\dagger$ & 5.77s\\
\hline
\end{tabular}
\end{center}
\label{table:comp}
\end{table}

\section{Conclusions}
In this paper, methods have been developed to compute distributed controllers that achieve an $\mathscr{H}_2$ performance bound for interconnected linear discrete-time systems with arbitrary interconnection structure. Convex controller existence conditions have been derived in the form of LMIs, which provide a scalable approach to the construction of distributed $\mathscr{H}_2$ controllers. Motivated by applications where communication between controllers is not possible, we have provided convex conditions for the existence of decentralized $\mathscr{H}_2$ controllers, through a suitable modification of the distributed $\mathscr{H}_2$ conditions. We have observed a considerable reduction in computation time with respect to centralized $\mathscr{H}_2$ controller synthesis for moderately-sized networks and efficient computation for large-scale networks for which the centralized $\mathscr{H}_2$ synthesis is not tractable.

\appendix
\section{Closed-loop matrices in Corollary \ref{cor:cl}} \label{app:clscales}
\begingroup
\allowdisplaybreaks
\begin{align*}
(Z_i^{11})_\mathcal{P}&:=-\operatorname*{diag}_{j\in\mathbb{Z}_{[1:L]}} (X_{ij}^{11})_\mathcal{P}, (Z_i^{22})_\mathcal{P}:=\operatorname*{diag}_{j\in\mathbb{Z}_{[1:L]}} (X_{ji}^{11})_\mathcal{P},\\
(Z_i^{12})_\mathcal{P}&:= \operatorname{diag}\left( -\operatorname*{diag}_{j\in\mathbb{Z}_{[1:i]}} (X_{ij}^{12})_\mathcal{P},\operatorname*{diag}_{j\in\mathbb{Z}_{[i+1:L]}} (X_{ji}^{12})_\mathcal{P}^\top \right),\\
(Z_i^{11})_\mathcal{C}&:=-\operatorname*{diag}_{j\in\mathbb{Z}_{[1:L]}} (X_{ij}^{11})_\mathcal{C}, (Z_i^{22})_\mathcal{C}:=\operatorname*{diag}_{j\in\mathbb{Z}_{[1:L]}} (X_{ji}^{11})_\mathcal{C},\\
(Z_i^{12})_\mathcal{C}&:= \operatorname{diag}\left( -\operatorname*{diag}_{j\in\mathbb{Z}_{[1:i]}} (X_{ij}^{12})_\mathcal{C},\operatorname*{diag}_{j\in\mathbb{Z}_{[i+1:L]}} (X_{ji}^{12})_\mathcal{C}^\top \right),\\
(Z_i^{11})_\mathcal{PC}&:=-\operatorname*{diag}_{j\in\mathbb{Z}_{[1:L]}} (X_{ij}^{11})_\mathcal{PC}, (Z_i^{22})_\mathcal{PC}:=\operatorname*{diag}_{j\in\mathbb{Z}_{[1:L]}} (X_{ji}^{11})_\mathcal{PC},\\
(Z_i^{12})_\mathcal{PC}&:= \operatorname{diag}\left( -\operatorname*{diag}_{j\in\mathbb{Z}_{[1:i]}} (X_{ij}^{12})_\mathcal{PC},\operatorname*{diag}_{j\in\mathbb{Z}_{[i+1:L]}} (X_{ji}^{12})_\mathcal{CP}^\top \right),\\
(Z_i^{12})_\mathcal{CP}&:= \operatorname{diag}\left( -\operatorname*{diag}_{j\in\mathbb{Z}_{[1:i]}} (X_{ij}^{12})_\mathcal{CP},\operatorname*{diag}_{j\in\mathbb{Z}_{[i+1:L]}} (X_{ji}^{12})_\mathcal{PC}^\top \right).
\end{align*}
\small
\bibliographystyle{ifacconf}
\bibliography{../../rfrncswp.bib}
\end{document}